\documentclass{aastex}
\usepackage{emulateapj5}

\newcommand{\bov}{\mbox{\boldmath $v$}}
\newcommand{\bj}{\mbox{\boldmath $J$}}
\newcommand{\bb}{\mbox{\boldmath $B$}}
\newcommand{\bnabla}{\mbox{\boldmath $\nabla$}}

\def\km{{\rm\,km}}

\def\mm{{\rm\,mm}}

\def\mum{\,\mu{\rm m}}
\def\K{{\rm\,K}}

\def\AU{{\rm\, AU}}

\shorttitle{Chondrule Formation by Current Sheets}
\shortauthors{Joung, Mac Low, \& Ebel}

\begin{document}

\title{Chondrule Formation and Protoplanetary Disk Heating by Current Sheets in Non-Ideal Magnetohydrodynamic Turbulence}

\author{M.~K.~Ryan~Joung\altaffilmark{1,2}, Mordecai-Mark~Mac~Low\altaffilmark{2,1}, and Denton~S.~Ebel\altaffilmark{3}}
\altaffiltext{1}{Department of Astronomy, Columbia University, 550 West 120th Street, New York, NY~10027; moo@astro.columbia.edu}
\altaffiltext{2}{Department of Astrophysics, American Museum of Natural History, 79th Street at Central Park West, New York, NY~10024; mordecai@amnh.org}
\altaffiltext{3}{Department of Earth and Planetary Sciences,
American Museum of Natural History, 79th Street at Central Park West, New York, NY~10024; debel@amnh.org}

\begin{abstract}
We study magnetic field steepening due to ambipolar diffusion \citep{bra94} in protoplanetary 
disk environments and draw the following conclusions. Current sheets are generated in 
magnetically active regions of the disk where the ionization fraction is high enough for 
the magnetorotational instability to operate. In late stages of solar nebula evolution, the 
surface density is expected to have lowered and dust grains to have gravitationally settled 
to the midplane. If the local dust-to-gas mass ratio near the midplane is increased above 
cosmic abundances by factors $\gtrsim\!10^3$, current sheets reach high enough temperatures to 
melt millimeter-sized dust grains, and hence may provide the mechanism to form meteoritic 
chondrules. In addition, these current sheets possibly explain the near-infrared excesses 
observed in spectral energy distributions (SEDs) of young stellar objects. Direct imaging 
of protoplanetary disks via a nulling interferometer or, in the future, a multi-band, adaptive 
optics coronagraph can test this hypothesis.
\end{abstract}


\section{Introduction}

Chondrules are millimeter-sized silicate spheres with igneous textures embedded in the most common type of meteorites, comprising 50--80 \% of their mass. Their formation is a long-standing mystery in solar system formation. Radioisotope datings place them among the first solids to have condensed out of the slowly cooling solar nebula. As such, they represent the remains of early materials that later formed the terrestrial planets. Their nearly spherical shapes and glassy compositions indicate that they cooled within a few hours after being molten, but not in the few seconds that objects that size would cool in free space. The associated heating event, whatever the origin, must have been transient and localized as evidenced by the inferred rapid heating and cooling rates, and the observed retention of volatiles such as sulfur (which vaporizes by prolonged heating above 650 K). To explain the high abundance of chondrules, it must have been a widespread phenomenon. Hence, encrypted in chondrules are important clues to physical conditions that prevailed in the early stages of solar system formation. Some even argue that taking the form of millimeter-sized molten spheres is a necessary intermediate step for initially micron-sized dust grains to grow to kilometer-sized objects \citep{hew97}.

A successful theory of chondrule formation should at least meet the following constraints: (1) the extremely short heating timescale of roughly seconds to minutes \citep{con98}, (2) the cooling rate of $10^2$--$10^3$ K hr$^{-1}$ \citep{rad90,lof90}, and (3) the narrow range of sizes (0.1--1$\mm$), inferred from laboratory analyses of textures and compositions as well as from attempts to reproduce chondrules. In addition, many chondrules show evidences of multiple heating events, so the heating event must have occurred more than once. Since the initial discovery 200 years ago \citep{how02}, their formation process has received considerable attention in both meteoritics and solar system astronomy, yet it remains elusive \citep{woo96}. Proposed mechanisms in the past \citep[see the summary by][for the pros and cons]{bos96} include nebula shocks, lightning, and an X-wind. None of these mechanisms have really been widely accepted \citep{hew96}, although recent work on shock heating by \citet{des02} has gained wide attention. 

Infrared observations and subsequent theoretical modeling of T Tauri stars revealed the existence of circumstellar (protoplanetary) disks \citep{men68,shu87,ada87,bec90}, many of which are actively accreting \citep[see, e.g. the review by][]{cal00}. These systems are generally thought to be analogs of the early solar nebula. On the theoretical side, magnetorotational instability (MRI) provided a solution to the long-standing problem of identifying the cause of anomalous viscosity, which allows mass accretion of the disk \citep[collectively referred to as BH91 hereafter]{bal91,haw91}. When applied to partially ionized disks, it remains an efficient mechanism of generating turbulence as long as ions are sufficiently well-coupled to neutrals, i.e., if ion-neutral collision rate far exceeds the orbital rotation rate: $\nu_{in} \gg \Omega$ \citep{bla94,mac95,haw98}. In the case of protoplanetary disks, however, a large portion, especially near the disk midplane, is associated with relatively low ionization fraction ($\sim 10^{-12}$), and hence thought to be magnetically dead \citep{gam96,gla97}.

In partially ionized disks, ambipolar diffusion arises because magnetic forces act on the charged component, i.e. ions, only \citep{mes56}. As the ions drift through neutrals, the magnetic field lines drag the ions with them, steepening field gradients near magnetic nulls stretched by shear flows \citep[hereafter BZ94]{bra94}. Consequently, sheets of high electric current form, as was confirmed by numerical simulations \citep{mac95,suz96}. BZ94 originally proposed this as a mechanism for rapid magnetic reconnection in the interstellar medium. Ambipolar diffusion is important in another context: as a heating source \citep{sca77,gol78,zwe83,elm85}. Recently, using three-dimensional simulations of turbulent, magnetized molecular clouds, \citet{pad00} pointed out that ambipolar diffusion heating is a significant source of heating in molecular clouds and that it can be larger than the cosmic ray heating. In this paper, we examine the possibility that similar physics operates in protoplanetary disks, such as in the solar nebula that eventually became our solar system.

The purpose of this paper is twofold. First, we show that current sheets are generated in magnetically active regions within protoplanetary disks. We study the properties of these current sheets and determine when and where the associated heating becomes significant. Second, we apply this formalism to the following two unresolved issues: the heating mechanism that melted chondrules in the early solar nebula and the near-infrared excesses observed in spectral energy distributions (SEDs) of young stellar objects. 

We later demonstrate that the heating of dust grains by current sheets is
ineffective for a standard model of the early solar nebula in which
the dust-to-gas ratio $\zeta \simeq 0.01$. However, at lower gas densities and 
much higher dust-to-gas ratios than those usually assumed
in the standard model, these current sheets are found to reach high
enough temperatures to melt millimeter-sized dust grains. We propose
that these unusual conditions occur near the midplane
in late stages of solar nebula evolution, when the gas surface
density has decreased due to various mechanisms
discussed below, and dust grains have gravitationally settled. The surfaces of 
this dust-enriched midplane layer are
then sufficiently dense and ionized to host thin ($l \approx
10^2\km$), hot ($T > 1600\K$) current sheets. Circulation of dust to the 
surface by turbulence within the dust layer is invoked to explain the observed
high abundance of chondrules.

The basic features of our current sheet model are set forth in \S\ \ref{csmodel}. In \S\ \ref{energy} we use this model to investigate a scenario in which chondrules are heated inside current sheets, solving numerically the coupled equations of energy transfer for gas and dust particles. The results are presented in \S\ \ref{results}. We argue, in \S\ \ref{midplane}, that in late stages of solar nebula evolution current sheets capable of melting most of the millimeter-sized dust particles form in the dust-rich midplane. In \S\ \ref{nearir} we discuss possible observational signatures of current sheets. Finally, in \S\ \ref{caveats} we list limitations of our model.

\section{Current Sheet Model}
\label{csmodel}
\subsection{Basic Equations}
\label{eqns}
 
The magnetic field evolves according to the induction equation,
\begin{equation}
\label{ind1}
\frac{\partial \bb}{\partial t} = \bnabla \times (\bov_e \times \bb - \eta \bnabla \times \bb) \, ,
\end{equation}
where $\bov_e$ denotes the electron velocity and $\eta$ is the associated resistivity given by 
\begin{equation}
\label{eta}
\eta = \frac{c^2}{4\pi \sigma_e} = \frac{234}{\chi_e} \, T_g^{1/2} \; \rm{cm^2 \, s^{-1}} \,
\end{equation}
\citep{kra73}. Here, $c$ is the speed of light, $\chi_e = n_e/n_n$ is the electron fraction, and $T_g$ is the ambient gas temperature. For the last equality in Eqn. (\ref{eta}), we adopt the value of the electrical conductivity, $\sigma_e$, from \citet{dra83} and \citet{bla94}. Subscripts $e$, $i$, and $n$ denote quantities pertaining to electron, ion, and neutral components, respectively. Other quantities of relevance in this paper are the mean mass per particle of $m_n = 2.33 \, m_p$, where $m_p$ is the proton mass, and the mean ion mass of $m_i = 39 \, m_p$ at densities where ionized potassium, K$^+$, is generally the dominant ion species \citep{bal01}. 

In the ideal MHD limit, the electron velocity is indistinguishable from the center-of-mass velocity of the fluid. This approximation breaks down in two ways as $\chi_e$ decreases. First, ions slip through neutrals, and for low enough $\chi_e$, when ion pressure and inertia can be neglected, the strong coupling approximation can be used to compute the ambipolar drift velocity by balancing the Lorentz force $\bj \times \bb /c$ acting on the ions against the frictional drag $\rho_i \nu_{in} (\bov_i - \bov_n)$ arising from collisions with neutrals. Second, when ions and electrons move along a direction perpendicular to a magnetic field, they experience the Lorentz force acting normal to both directions but they drift away in two opposite directions due to the different signs of their electric charges. This causes the Hall effect. The electron velocity in a weakly ionized medium is then
\begin{equation}
\label{ve}
\bov_e = \bov + (\bov_i - \bov) + (\bov_e - \bov_i) = \bov + \frac{\bj \times \bb}{c \rho_i \nu_{in}} - \frac{\bj}{n_e e} \, ,
\end{equation}
where $\rho_i = m_i n_i$ and $\nu_{in} = \gamma \rho_n$ are the ion mass density and ion-neutral collision rate. The value of the drag coefficient, $\gamma = 2.75 \times 10^{13}$ cm$^3$ s$^{-1}$ g$^{-1}$ \citep{bal01}. Since $\chi_e \ll 1$ in our case (see \S\ \ref{mri}), the neutral velocity is the same as the center-of-mass velocity, $\bov$. The current density has the usual form $\bj = (c/4\pi) \bnabla \times \bb$. We plug Eqn. (\ref{ve}) into Eqn. (\ref{ind1}) and consider the case in which the currents and fields are perpendicular ($\bj \cdot \bb = 0$), so the induction equation becomes
\begin{equation}
\label{ind2}
\frac{\partial \bb}{\partial t} = \bnabla \times (\bov \times \bb) - \bnabla \times \left[ \left(\frac{B^2}{4\pi \rho_i \nu_{in}} + \eta \right) \frac{4\pi \bj}{c} \right] \, .
\end{equation}
The Hall term does not affect the time evolution of $\bb$, since $\bnabla \times (\bj \times \bb) \propto -\bnabla \times \bnabla B^2$ vanishes. Thus, ambipolar diffusion acts as \textit{an additional field-dependent resistivity}. We define the ambipolar diffusion coefficient as 
\begin{eqnarray}
\lambda_{AD} &\equiv& \frac{B^2}{4\pi \rho_i \nu_{in}} \\
&\approx& 1.1 \times 10^{17} \, B_G^2 \left( \frac{n_n}{10^{13} \; \rm{cm^{-3}}} \right)^{\! -2} \! \left( \frac{\chi_i}{10^{-12}} \right)^{\! -1} \, \rm{cm}^2 \, \rm{s}^{-1} \, , \nonumber
\end{eqnarray}
where $B_G$ means $B$ in gauss.

For our model of the protoplanetary disk, we take the minimum-mass solar nebula of \citet{hay81}. It is characterized by the surface mass density $\Sigma = \Sigma_0 \, (R/3 \AU)^{-3/2}$ with $\Sigma_0 = 3.3 \times 10^2$ g cm$^{-2}$, where $R$ is the radial distance from the central star, and the temperature profile given by $T_g= 1.6 \times 10^2 \, (R/3 \AU)^{-1/2}$. We expect $\Sigma_0$ to decrease in time as the nebular gas clears out in the late stage of protoplanetary disk evolution. Hence we use the form $\Sigma_0 = 3.3 \times 10^2 f_{\Sigma}$ g cm$^{-2}$, where $f_{\Sigma}$ is an arbitrary parameter less than or equal to unity. If we take the gas to be isothermal, the vertical density profile in hydrostatic equilibrium has a Gaussian form \citep[][hereafter CG97]{ken87,chi97}
\begin{equation}
n_g = n_0 \, \exp \left( -\frac{z^2}{2 h^2} \right) \, ,
\end{equation}
where $n_0 = \Sigma / (\sqrt{2\pi} \, m_n h) = 1.4 \times 10^{13} \, f_{\Sigma} \, (R/3 \AU)^{-11/4}$ cm$^{-3}$ is the number density of gas in the disk midplane, $z$ is the vertical distance measured from the midplane, and $h \equiv c_s/\Omega$ is the scale height. Here, as usual, $c_s = (k T_g/m_n)^{1/2}$ is the sound speed and $\Omega$ is the Keplerian rotation rate. With these values, we can roughly estimate the importance of ambipolar diffusion by taking the ratio of the two coefficients in Eqn. (\ref{ind2}):
\begin{eqnarray}
\label{ratio}
\frac{\lambda_{AD}}{\eta} &=& \frac{B^2}{4\pi \eta \rho_i \nu_{in}} \\
&\approx& 20 \, f_{\Sigma}^{-2} \, B_G^2 \left( \frac{R}{3 \AU} \right)^{\! 23/4} \! \exp \!\left( \frac{z^2}{h^2} \right) \, . \nonumber
\end{eqnarray}
We note that (1) with a magnetic field sufficiently amplified (up to $\sim$1 G), ambipolar diffusion is at least as important as Ohmic dissipation and (2) the ratio has strong spatial dependence within the disk. Remnant magnetism in meteorites suggests that $B \approx$ 1 G was typical in the solar nebula. If an efficient dynamo is in action, the ambipolar diffusion term would generally be important in the outer ($R \gtrsim$ 1 AU) region of the disk. In the inner region, for a similar value of $B$ and $f_{\Sigma}=1$, it is nonnegligible only a few scale heights above and below the midplane, where gas density is much lower. It generally becomes increasingly important as $f_{\Sigma}$ decreases.

\subsection{Magnetorotational Instability in Protoplanetary Disks}
\label{mri}

Compared to other astrophysical systems, protoplanetary disks are unusually cold and dense and are therefore associated with very low ionization fractions. Ionization fraction is a key parameter in determining whether a disk will be turbulent or not, since the MRI is confined to regions where the ionization fraction is high enough to ensure good coupling between ions and neutrals. This occurs when the ion-neutral collision rate ($\nu_{in}$) is significantly greater (by 3--100 at least) than the rotation rate ($\Omega$) \citep{bla94,mac95,haw98}. The critical ionization fraction is calculated to be
\begin{equation}
\chi_{crit}\! = \!\frac{\Omega}{m_i n_n \gamma}\! = 2 \times 10^{-12} f_{\Sigma}^{-1} \!\left( \frac{R}{3 \AU} \right)^{\! 5/4} \!\! \exp\!\left( \frac{z^2}{2 h^2} \right)
\end{equation}
\citep[e.g.,][]{bal98}. Interestingly, this value is only slightly higher than the values of $10^{-12}$--$10^{-13}$ estimated in the bulk of protoplanetary disks \citep{gam96,gla97,san00}. 

Many numerical simulations have demonstrated that MRI-induced velocity shears amplify the magnetic field \citep[see, e.g.,][]{haw91,brae95,haw95}. In our model, perturbations in the magnetic field are also assumed to be generated by the MRI. Considering the low values of $\chi_i$ estimated in protoplanetary disks, it is initially unclear whether this assumption is valid or not. Therefore we ask: which region of the solar nebula, if any, is prone to the MRI? (Previous numerical simulations assumed ionization fraction $\chi_i \gg \chi_{crit}$.) This question has been investigated by several authors \citep{ume83,san00,fro02}. Their results are as follows. The ionization fraction is sufficiently high in the innermost region of the disk up to a few tenths of an AU, where the ambient gas temperature is high enough ($T_g \gtrsim 10^3$ K) for collisional ionization to be effective \citep[where it dominates over cosmic ray and radioactive element ionizations;][]{ume83}. Also, cosmic rays and stellar X-ray radiation ionize the surface layers (with thickness given by the stopping depth of $\Sigma_a \approx 10^2$ g/cm$^2$) at $R \lesssim 10$ AU and the entire disk at $R \gtrsim 10$ AU, where the surface density of the disk falls below $\Sigma_a$. These, then, are the regions where the disk is turbulent, hence effectively transferring angular momentum and also actively accreting. On the other hand, the region near the disk midplane, where terrestrial planets and the asteroid belt now lie, is generally thought to contain no significant heating source and be quiescent, i.e. magnetically dead \citep[radioactive heating is too inefficient to maintain $\chi_i > \chi_{crit}$;  see][]{gam96}. These considerations led to the layered accretion model of Gammie \citep[1996; see also][]{gla97}.

There are two length scales, other than the scale height $h$, of relevance to this discussion. The wavelength of perturbation most unstable to the MRI is only slightly larger than the critical wavelength at which the instability sets in (BH91) and is given by
\begin{equation}
\lambda_{BH} = \!\frac{2\pi}{\sqrt{3}} \frac{v_A}{\Omega} \approx 0.23 \, f_{\Sigma}^{-1/2} \, B_G \!\left( \frac{R}{3 \AU} \right)^{\! 23/8} \!\! \exp \!\left( \frac{z^2}{4h^2} \right) \, \AU \, 
\end{equation}
(BH91). Other processes affect the growth of perturbations: in dense parts of the disk, collisions with neutrals hinder the charged components from drifting with the magnetic field. In this case, Ohmic dissipation acts to stabilize wavelengths shorter than the resistive diffusion scale $\lambda_{\Omega} \equiv 2 \pi \eta /v_A$, where $v_A=B/(4\pi \rho_n)^{1/2}$ is the Alfven speed. In other words, the critical wavelength at which MRI sets in is given by $\lambda_{crit} = max \, (\lambda_{BH},\lambda_{\Omega})$ \citep{bla94,san00}. When $\lambda_{\Omega} > \lambda_{BH}$, the perturbation growth becomes much slower. An additional constraint enters when $\lambda_{crit}$ becomes comparable to the scale height $h$; at this point, saturation of the MRI is thought to occur \citep{bal98}. 

\begin{figure}
\plottwo{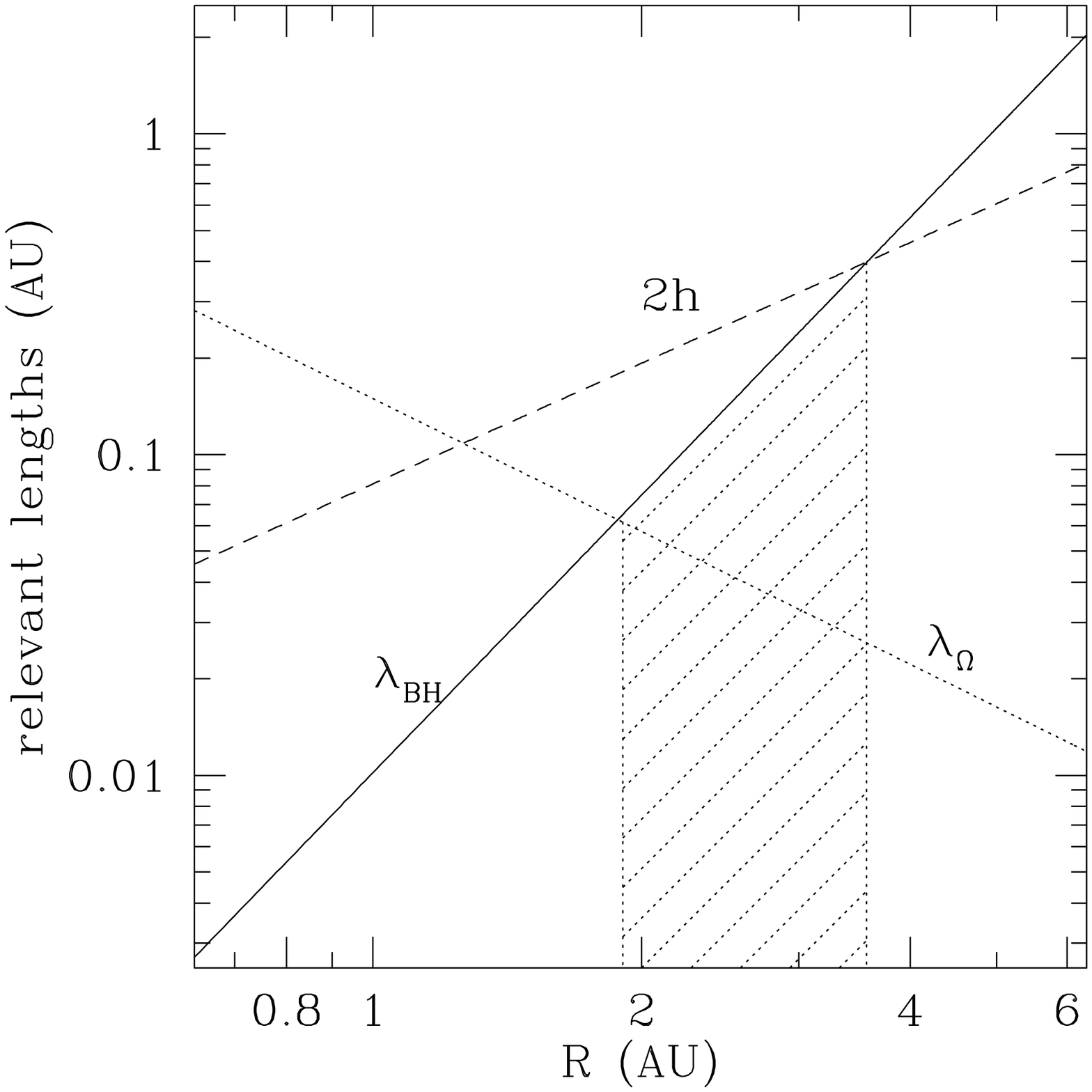}{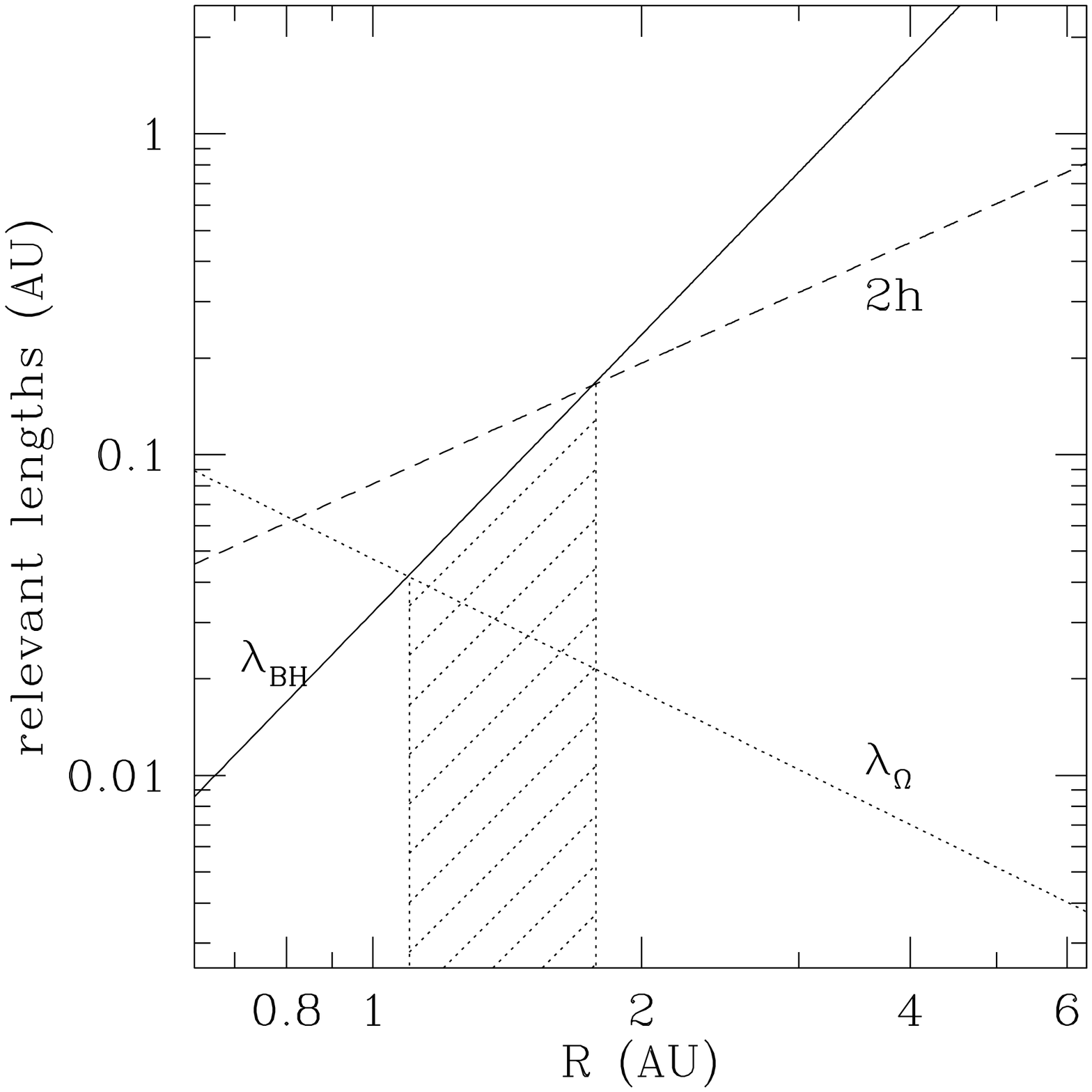}
\caption{Relevant length scales. (a) Assumed parameter values are $f_{\Sigma} = 1$, $B$ = 1 G, and $\chi_i = 10^{-12}$. The critical wavelength of the MRI, resistive diffusion scale, and (twice) the scale height of the disk are shown as solid, dotted, and dashed lines, respectively. See text for their definitions. The shaded region satisfying $\lambda_{\Omega} < \lambda_{BH} < 2h$, where the MRI occurs, extends from 2 AU to 4 AU. (b) Same as (a) but using $f_{\Sigma} = 10^{-1}$. This is appropriate for a late stage of the solar nebula evolution. Note that the magnetically active region moves inward to 1--2 AU. \label{fig1}}
\end{figure}

\citet{san00} investigated where the MRI operates in protoplanetary disks 
by including a detailed chemical reaction scheme and the effect of the recombination of 
ions and electrons on grain surfaces. They showed that for the minimum-mass solar nebula 
($f_{\Sigma}=1$) the 2--3 AU region of the disk would be stabilized by Ohmic dissipation. 
However, as the surface density parameter, $f_{\Sigma}$, is lowered, the unstable region 
expands (see their Figure 8). Also, as the disk evolves, dust grains settle to a thin 
layer in the midplane, and so the grain abundance in the bulk of the nebula gradually 
decreases. For this reason, Sano et al. (2000) introduced a 
dust depletion factor, $f_d$, which measures the abundance of dust grains relative to that 
in molecular clouds. Thus a low value of $f_d$ may correspond to an old, evolved disk. 
They demonstrated that as $f_d$ decreases for a given surface density, the stable region 
(the ``dead zone'') shrank in size, and for $f_d \lesssim 10^{-2}$, the entire disk becomes 
unstable for $R > 3$ AU (see their Figure 11). This happens because the recombination 
rate on grain surfaces decreases. They found a similar trend as grain 
growth occurs. From these results, we conclude that the region at $R \approx 3$ AU 
will become magnetically active at a sufficiently late stage of the evolution. We use and 
discuss this result further in \S\ \ref{apply}. Figure \ref{fig1} compares the three 
relevant length scales at an early (a) and a late stage (b) of solar nebula evolution. 
The region of interest where magnetic fields amplify due to the MRI extends from 2 AU to 
4 AU in (a), while with a reduced gas density, the active region moves inward to 1--2 
AU (b). 

\subsection{Formation of Current Sheets}
\label{formcs}

\begin{figure}
\epsscale{0.45}
\plotone{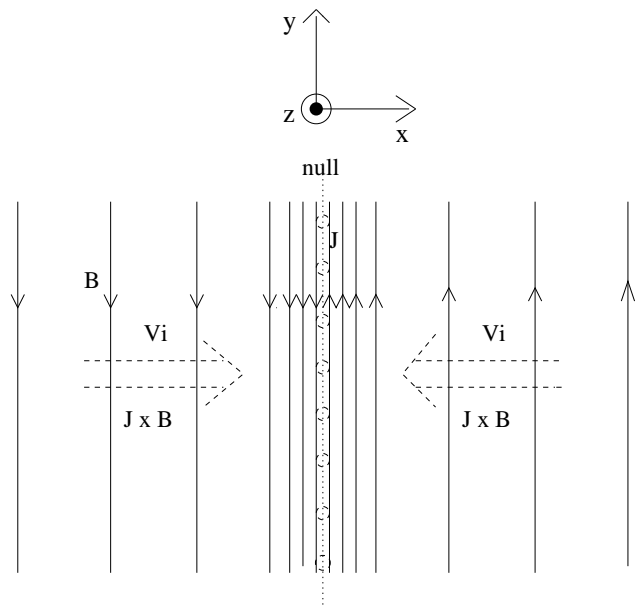}
\caption{Configuration for magnetic fields in the one-dimensional geometry considered in this paper. Current density ($J$) points in the $+z$ direction, out of the page. The directions of various terms in Eqns. (3) and (4) are also shown. A current sheet forms along the magnetic null, shown as a dotted vertical line along the center of the figure. \label{fig2}}
\end{figure}

In this and the following sections, we consider a simple, one-dimensional geometry illustrated in Figure \ref{fig2}. Also shown are the directions of various terms in Eqns. (\ref{ve}) and (\ref{ind2}). Consider first a case in which a magnetic null is present. Near the null, a magnetic pressure gradient, $-\bnabla (B^2/8\pi)$, acts on ions and pushes the field lines into the null point in both directions. The ion-neutral drift is then directed toward the null. Along a gradient in the magnetic field, this nonlinear term produces faster field diffusion in regions of stronger field, driving field into regions of weaker field and steepening the gradient, rather than spreading it out as the linear Ohmic term does. Ultimately, the gradient sharpens towards a singularity, marked by a sheet of high electric current. This steepening of magnetic field profile (BZ94) occurs within a timescale of $\tau_{AD} \equiv L_{BH}^2/(2 \, \lambda_{AD}) = (\pi^2/\,6) \, (\nu_{in}/\Omega^2)$, calculated to be $\pi/12$ of the orbital period when $\chi_i=\chi_{crit}$. The factor of 1/2 in $\tau_{AD}$ takes into account the fact that the characteristic length decreases as steepening proceeds. 

Remarkably, this sharpening has been shown to occur even in the absence of nulls due to a variety of processes including a rotating eddy \citep[BZ94;][]{zwe97}, accretion disks sustaining the MRI \citep{mac95}, large amplitude Alfven waves \citep{suz96}, and the nonlinear stage of Wardle C-shocks \citep{mac97}. As previously mentioned, we take the example of magnetohydrodynamic turbulence driven by the MRI. In this case, velocity shears produce magnetic nulls and current singularities . Stretched by shear flows, magnetic nulls develop where the magnetic field changes directions. We therefore crudely estimate that the separation between two adjacent magnetic nulls $L_{BH}=\lambda_{BH}/2$. 

We note that ambipolar diffusion allows the field to slip through the neutrals and tends to weaken the MRI. For strong enough diffusion, i.e. when the ion-neutral collision rate ($\nu_{in}$) is much higher than the growth rate of the MRI ($\sim\Omega$), the instability is entirely prevented \citep{mac95,haw98}. On the other hand, ambipolar diffusion is required to generate current sheets. For our model, we use parameters on the interface between the two regimes. This choice is justified by two-dimensional simulations performed by \citet{mac95}, which showed that for drag coefficients $\gamma$ a few times the minimum value that allows the instability to develop, $\gamma_{min} \equiv \Omega /(m_i \, n_n \, \chi_{crit})$, both the instability and ambipolar diffusion sharpening occur. In other words, we require $1 \lesssim \gamma/\gamma_{min} < 100$. Alternatively, magnetic field amplification could have preceeded the epoch of current sheet formation.

We now make quantitative estimates of the properties of current sheets. BZ94 showed that, in the absence of resistivity, the magnetic field profile relaxes to $B \propto x^{1/3}$, and the current density $J \propto \partial B/\partial x \propto x^{-2/3}$ becomes singular at the origin. In reality, both the finite resistivity and the ion pressure act to remove this singularity \citep{bra95}. Under the physical conditions of protoplanetary disks, ion pressure can be neglected \citep{hei03}. The induction term in Eqn. (\ref{ind2}), $\bnabla \times (\bov \times \bb)$, becomes progressively less important as the field steepens; $I/O \approx vL/\eta$ where $I$ and $O$ denote the magnitudes of the induction and Ohmic dissipation term respectively, and is therefore ignored. The quasi-steady-state magnetic field profile (``Quasi'' since the magnetic field profile lasts only as long as the velocity shear is present, for about a dynamical time) is calculated by balancing the Ohmic dissipation term against the ambipolar diffusion term in Eqn. (\ref{ind2}):
\begin{equation}
\label{balance}
(B_0^2+B_y^2) \frac{d^2 B_y}{d x^2} + 2 B_y \left( \frac{d B_y}{d x} \right)^{\!2} = 0 \, .
\end{equation}
Here, for simplicity, we use the definition
\begin{equation} 
\label{b0}
B_0 \equiv (4\pi \eta \rho_i \nu_{in})^{1/2} \approx 0.26 \, n_{g13} \, T_{g3}^{1/4} \; \rm{G} \, ,
\end{equation}
which measures the relative importance of Ohmic dissipation to ambipolar diffusion. We used $n_{g13}$ and $T_{g3}$ to denote $n_g/(10^{13} \; \rm{cm^{-3}})$ and $T_g/(10^3 \; \rm{K})$, respectively. Ohmic dissipation dominates where $B < B_0$; ambipolar diffusion wins where $B > B_0$ (see Eqn. \ref{ratio}). Remarkably, $B_0$ is independent of the ionization fraction; so is the magnetic field profile. Solving Eqn. (\ref{balance}), we obtain
\begin{equation}
\label{cubic}
\frac{1}{3} B_y^3 + B_0^2 B_y = \Gamma x \, ,
\end{equation}
assuming that the field vanishes at the origin. An integration constant, $\Gamma$, is determined by the boundary condition $B_y(x=L_{BH})=B_{max}$, which leads to $\Gamma = B_{max}(B_0^2+B_{max}^2/3)/L_{BH}$. The amplitude of the magnetic field, $B_{max}$, is determined from the remnant magnetic field strengths recorded in the matrix of chondrites (0.1--1 G) and in chondrules themselves (1--10 G) \citep[][and references therein]{des02}. From Eqn. (\ref{cubic}), we see that $B_0^2 \gg B_y^2 /3$ near the origin, and the effect of resistivity yields $B_y \approx \Gamma x /B_0^2$. Note the current density singularity is now removed; the maximum current density near the origin is given by $J(x=0) = c \Gamma / 4\pi B_0^2$, which is finite. Sufficiently far from the origin where $B_0^2 \ll B_y^2 /3$ is satisfied, $B_y \approx (3 \Gamma x)^{1/3}$ holds, and we retain the result in the non-resistive limit of BZ94. There, the current density declines steeply ($J \propto x^{-2/3}$). We therefore set $B_0^2=B_y^2/3$ and compute the width of the current sheet to be
\begin{eqnarray}
\label{lcs}
l_{cs} &=& \frac{2 \sqrt{3} B_0^3}{\Gamma} \\
&\approx& 1.5 \times 10^5 \, n_{g13}^{5/2} \, T_{g3}^{3/4} \, B_G \left( \frac{B_{max}}{3 \; \rm{G}} \right)^{\!-3} \! \left( \frac{R}{3 \AU} \right)^{\!3/2} \; \rm{km} \, . \nonumber
\end{eqnarray}
This expression is valid for $B_{max} > \sqrt{3} \, B_0$, which is satisfied under the conditions of interest.

\subsection{Heating Mechanisms}
\label{hmech}

Inside current sheets, two heating mechanisms act. First, electrons and ions are accelerated by the electric field present, and then collide with neutrals and heat up the entire gas, leading to familiar Ohmic dissipation. The corresponding heating rate per unit volume essentially follows from the magnetic field profile:
\begin{equation}
\varepsilon_{\Omega} = \frac{j^2}{\sigma_e} = \frac{\eta}{4\pi} \left( \frac{d B_y}{d x} \right)^{\!2} \, .
\end{equation}
There is additional heating due to friction when ions, accelerated by the Lorentz force, diffuse through neutrals \citep[see, e.g., \S\ \ref{eqns} of this paper; ][]{pad00}. In this case of ambipolar diffusion heating,
\begin{equation}
\varepsilon_{AD} = \rho_i \, \nu_{in} | \bov_i - \bov_n |^2 = \frac{\eta B_y^2}{\pi B_0^2} \left( \frac{d B_y}{d x} \right)^{\!2} \, 
\end{equation}
yields the volumetric heating rate. Note that their ratio is given simply by $\varepsilon_{AD}/\varepsilon_{\Omega}=B_y^2/4B_0^2$. Within the range of parameters we are interested in, the ambipolar diffusion heating integrated through the current sheet along $x$ exceeds (by a factor of $\sim$10 in our fiducial model) the contribution from Ohmic dissipation.

\begin{figure}
\epsscale{1.0}
\plottwo{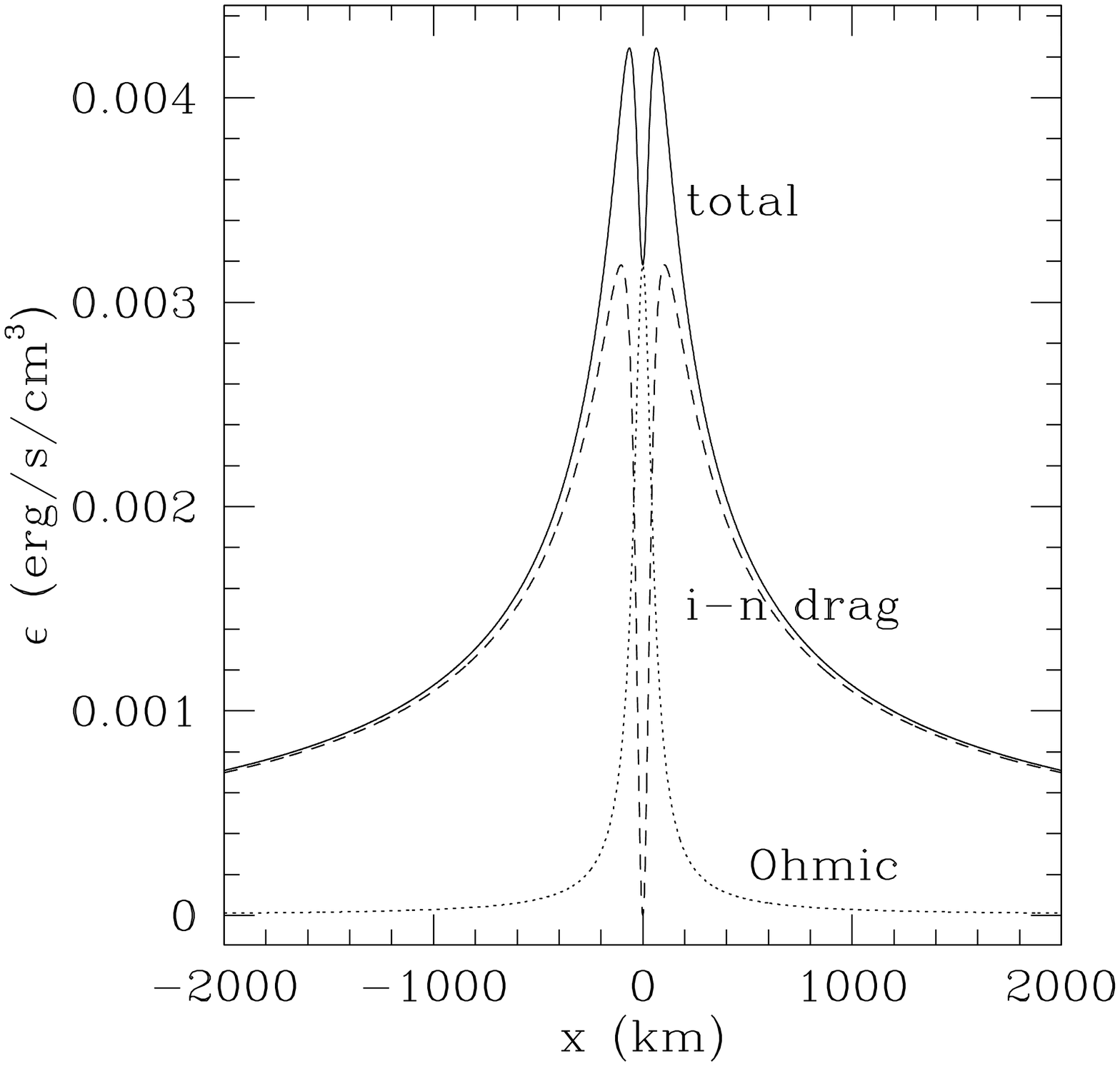}{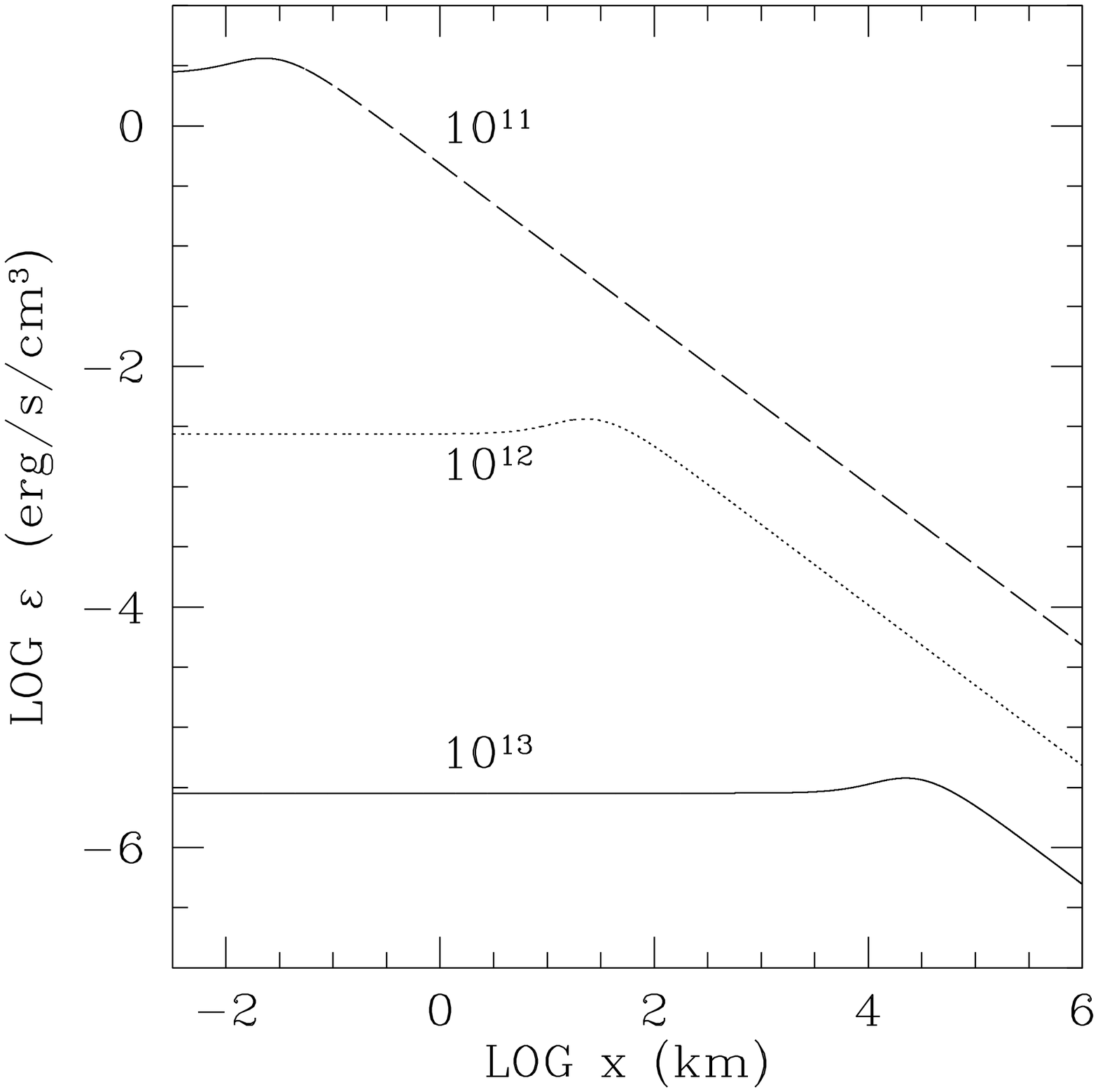}
\caption{(a) Heating profile for our fiducial model with $f_{\Sigma} = 0.1$. Note the symmetry about $x=0$. Ohmic dissipation (dotted line) is concentrated near center of current sheet, while the ambipolar diffusion heating (dashed line) is centered off-peak. The latter exceeds the contribution from the former by a factor of $\sim$6 in our fiducial model. (b) Heating profiles as a function of $B_0$ in a log-log scale. In this case, the gas density ($n_g$) was varied, and each curve is labeled with the corresponding value of $n_g$ used. The peak heating rate increases approximately inversely proportional to the square of gas density; the width of current sheet varies as described by Eqn. (\ref{lcs}). \label{fig3}}
\end{figure}

The resulting heating profiles (log $\varepsilon$ vs. log $x$) as a function of $B_0$, defined in Eqn. (\ref{b0}), are displayed in Figure \ref{fig3}. They are naturally symmetric around $x=0$. In terms of shape, these double peaks are similar to what has been observed in numerical simulations of \citet{pad00}, although their sheets are produced by supersonic shocks and density compressions. Ambipolar diffusion sharpening becomes more efficient at lower gas densities (see Eqn. \ref{lcs}), and the current densities become close to singular. This implies stronger and more concentrated heating far above and below the disk midplane. Alternatively, if the disk becomes thinner with time \citep{rud99}, the heating rate also increases. For example, soon after thermonuclear reactions began in the Sun, a powerful solar wind is thought to have blown away the gas $\sim$0.1 to a few million years after the solar nebula formed. Accretion onto the central star or, in local regions, planetary accretion as gap formation occurs would have the same effect. A density drop caused by any of these mechanisms introduces a regime where chondrule formation is favored. As the overall gas density $n_g$ drops in time, $B_0 \propto n_g$, so the peak heating rate rises. 

\section{Applications}
\label{apply}
\subsection{Energy Equations and Chondrule Formation}
\label{energy}

We investigate chondrule heating in current sheets by  
solving numerically the coupled equations of energy transfer for gas and dust 
particles, and compare the results with the experimental constraints 
on chondrule formation.

Dust particles are usually assumed to possess a power-law distribution of sizes with upper 
and lower size cutoffs 
\citep[e.g.][]{chi01}. For simplicity, we use just two types of dust particles: 
small, micron-sized dust grains with radius $a_d=1\mum$ and bigger, millimeter-sized chondrules 
with $a_c=1\mm$. Also, it 
is assumed that the same amount of mass is contained in each type. Near the dust 
melting temperature, small grains 
radiate in the near-infrared most efficiently. Hence their thermal evolution is 
of considerable importance to the overall development of current sheets. 

To study how these current sheets affect the temperatures of dust particles that emit 
the bulk of the infrared radiation, we use the one-dimensional slab geometry 
considered in \citet{hoo93}. Dust and gas travel through a current 
sheet with relative velocity of $v_r$, where $v_r$ is a free parameter of the model 
but assumed to be comparable to the Alfven speed, $v_A=B/(4\pi \rho_n)^{1/2}$, and 
smaller than 17 km s$^{-1}$, the Keplerian orbital speed at $R=3$ AU. For our fiducial model, 
$v_r=1$ km s$^{-1}$. We discretize the region for computation so that each 
zone has a thickness of 1 km and we use $10^3$ zones. Note that the lateral extent 
of our computational domain exceeds that of our fiducial current sheet ($l_{cs} \approx 
10^2$ km). The gas and dust 
temperatures are uniform in each zone but in general differ from each other. The 
temperatures evolve in time and are assumed to vary only in one spatial direction 
($x$).

Entering a current sheet, gas molecules gain energy from the two heating mechanisms 
mentioned in \S\ \ref{hmech} and lose it through collisions with dust grains and 
chondrules. The 
collisional heat transfer rate per unit surface area between gas and dust, 
denoted by $q_d$, is defined as $q_d = (3/4\sqrt{\pi}) \, (2k/m_n)^{\! 3/2} \rho_g 
(T_g - T_d)$ and $q_c$ is defined similarly for chondrules \citep{hoo93}. Then the 
gas temperature evolves in time according to 
\begin{equation}
\label{tg}
n_g k \, \frac{d \, T_g}{d \, t} = \varepsilon - \sum_{i=d,c} 4\pi a_i^2 \, 
n_i \, q_i + \rho_g \kappa_{IR} \, \left( J_r - 4 \sigma T_g^4  \right) \, ,
\end{equation}
where $k$ is the Boltzmann constant, $n_i$ is the number density of either dust 
grains ($i=d$) or chondrules ($i=c$), and $\varepsilon = \varepsilon_{\Omega} + 
\varepsilon_{AD}$. The opacity of the gas, $\kappa_{IR}$, is mainly due to dust 
associated with it, and is set to 1.0 cm$^2$ g$^{-1}$, appropriate for infrared 
wavelengths \citep{chi01}. The intensity of radiation, $J_r$ (integrated over 
all wavelengths) at $x=x_0$ from plane-parallel, temperature-stratified atmospheres 
has the following form \citep{mih78,des02}:
\begin{equation}
\label{jr}
J_r (x_0) = \! 2 \sigma \! \int_{-\infty}^{+\infty} \!\!\! ( \rho_g \kappa_{IR} \, T_g^4 + \sum_{i=d,c} 
n_i \, \pi a_i^2 \, \epsilon_i T_i^4 ) \, E_1 (\beta |x-x_0|) \, dx \, 
\end{equation}
with $\beta \equiv \rho_g \kappa_{IR} \, T_g^4 + \sum_{i=d,c} n_i \, \pi a_i^2 \, \epsilon_i$. 
Here, $\epsilon_i$ is the emissivity of a grain of type $i$ averaged over the Planck 
function at temperature $T_i (x)$, which we simply set at 1.0 (for comparison, 
\citet{des02} uses values of 0.8 and 0.4 for 1$\mm$- and 1$\mum$-sized grains, 
respectively), and $E_1$ is the 
exponential integral of the first type. Dust grains and chondrules, on the other hand, 
are heated both by colliding with hotter gas molecules and by absorbing radiation 
from nearby smaller, micron-sized, grains which reach high temperatures quickly. 
They lose energy by radiating an approximate blackbody spectrum at $T_d$ and $T_c$, 
with emission efficiency of $\epsilon_d$ and $\epsilon_c$ for dust and chondrules, 
respectively. Therefore, 
\begin{equation}
\label{td}
m_i \, C_i \, \frac{d \, T_i}{d \, t} = \pi a_i^2 
\left[ 4 q_i + \epsilon_i \left( J_r - 4 \sigma T_i^4 \right) \right] \, .
\end{equation}
As before, the index $i$ can be either $d$ (for dust) or $c$ (for chondrule). The 
last term in Eqn. (\ref{td}) represents the incident flux on a given particle 
radiated by dust grains in the other zones minus its own emission.

The experimental constraints on chondrule formation are 
(1) the grains are completely melted for a few minutes; (2) the cooling 
timescale of gas and dust is approximately an hour, consistent with the oft-quoted 
value of $10^2$--$10^3$ K hr$^{-1}$; and (3) small grains ($\lesssim$ 0.1 mm) are 
absent, possibly evaporated or destroyed completely 
after passing through a current sheet, leading to the 
observed narrow size distribution centered at $a_d \approx$ 1 mm.

From the properties of current sheets described in \S\ \ref{formcs}--\ref{hmech}, 
we infer that the above constraints on chondrule formation will be satisfied only 
if the following conditions are met:

\noindent(1) In order to obtain high enough heating rate, the gas density has to be
low ($n_g \lesssim 10^{12}$ cm$^{-3}$), because the heating rate increases steeply 
as $n_g$ decreases.

\noindent(2) To have "localized" heating events, spatially averaged chondrule/dust
densities should be high ($n_d > 1$ cm$^{-3}$). Otherwise, the mean free path
for IR photons will exceed a few hundred kilometers and the heating timescale will
be much more than ``a few minutes.''

\noindent(3) To ensure good thermal coupling between gas and dust, either
the gas density has to be high ($n_g \gtrsim 10^{12}$ cm$^{-3}$) because 
$q_d$, $q_c \propto n_g$ which cannot be satisfied because of the first condition,
or dust grains need to be very small ($a_d \lesssim 1 \mum$) so they approach
$T_g$ quickly.

In order to satisfy conditions (1)--(3) simultaneously, the dust-to-gas mass ratio, 
$\zeta$, should be very large, about 50. Usually $\zeta \approx 0.01$ is
assumed for the minimum mass solar nebula, but after a few million years
the gas density is expected to have decreased due to viscous diffusion
\citep{rud86} as well as the various mechanisms mentioned in the last paragraph of 
\S\ \ref{hmech}. Dust settles into 
a thin, dense layer on a timescale $\tau_s \approx (a/1 \mum)^{-1} f_{\Sigma}$ Myr, 
independent of $R$  
\citep[][CG97]{rud99}, comparable to the lifetimes of protoplanetary disks. There will 
still be some dust grains, particularly submicron-sized ones, that remain well above the 
midplane mixed with the gas. But at late times the dust density outside the thin, 
dense dust layer is likely to be quite small, decreasing the dust-to-gas ratio 
there. Inside the dust layer, however, solids may dominate the 
local density \citep{cuz93}. Hence, $\zeta \approx 50$ may be reached in the 
midplane, where most solids reside. 

We solve Eqns. (\ref{tg})--(\ref{td}) simultaneously for $T_g(x, t)$, $T_d(x, t)$, 
and $T_c(x, t)$ using a fifth-order Runge-Kutta 
scheme appropriate for a stiff set of equations with adaptive stepsize control 
\citep{pre92}. Both gas and dust particles begin at the same temperature, $T_{init} 
= 500$ K. The center of a current sheet is initially placed at $x=1000$ km and 
moves to the left at $v_r$. To simplify the problem, we ignore the 
time evolution of $\varepsilon$ and $\chi_i$ (see \S\ \ref{caveats}). For our 
fiducial model, $n_g = 10^{12}$ cm$^{-3}$ and both $n_d$ and $n_c$ are set
to yield desired values of the dust-to-gas ratio $\zeta$.

\subsection{Results}
\label{results}

\begin{figure}
\epsscale{0.48}
\plotone{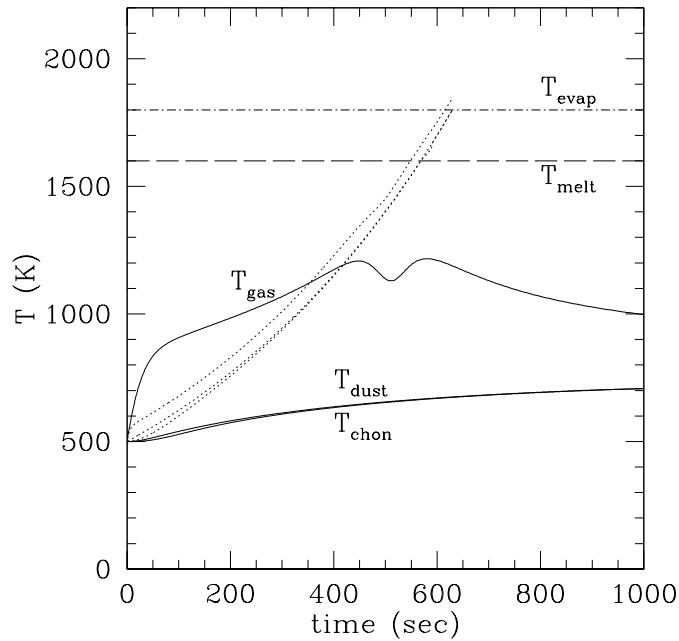}
\caption{Time evolution of gas, dust, and chondrule temperatures at a fixed position in space. 
Common parameters used for the 
two runs are: $n_g$ = 10$^{12}$ cm$^{-3}$, $a_d$ = 1 $\mu$m, $a_c$ = 1 mm, and 
$B_{max}$ = 3 G. Solid lines are used for the run with $\zeta = 5$ and dotted lines 
for the run with $\zeta=50$. Only for the higher value of $\zeta$ do dust grains and chondrules reach the melting temperature. We stop running the code when the particles evaporate. The initially steep rise in $T_{gas}$ is due to intial conditions that are not set self-consistently. \label{fig4}}
\end{figure}

Figure \ref{fig4} shows the time evolution of gas, dust, and chondrule 
temperatures at a fixed position as a current sheet moves through, for our model 
runs with $\zeta=$ (5, 50). We also 
ran the code for $\zeta=0.05$ but do not plot the result because both $T_d$ and $T_c$ 
remain practically unchanged at $T_{init}$ for over $10^3$ s. We find that 
for $\zeta = 5$ the gas temperature 
rises appreciably (above 1000 K) but, due to poor thermal coupling between gas 
and dust, neither $T_d$ nor $T_c$ changes significantly, reaching only $\sim 700$ 
K after 10$^3$ s. In the run with $\zeta=$ 50, however, due to the increase 
in optical depth, dust grains 
follow the gas temperature closely and the radiation field also steadily increases in 
strength. The grains melt after about 570 seconds and evaporate after about 630 seconds, 
at which point we stop the simulation, as it does not take into account the evaporation 
of particles. We 
changed the boundary condition between flat at $T_{init}$ and a floating boundary condition 
but found the result remained qualitatively the same. In an improved model that accounts 
for evaporation and a range of dust grain sizes, we expect the smallest grains 
to evaporate first, decreasing the optical depth, thus widening the heated region and 
lowering the net volumetric heating rate. The larger grains will therefore probably 
not evaporate. This is interesting since it could naturally explain the peaking of 
the size distribution of chondrules around 1 mm.

The fact that chondrules get melted only for high values of $\zeta$ is consistent  
with the FeO record in chondrules, which seems to indicate oxygen fugacities (relative 
abundance of oxygen to hydrogen) at formation that are higher than solar values by 
at least two or 
three orders of magnitude \citep{woo67}. This is most often interpreted to mean that 
chondrules were heated in regions with extremely high dust-to-gas ratios. The dust is 
taken to be oxides of Si, Mg, Ca, Al, etc. in chondritic proportions which bring in a 
large amount of oxygen. For 
this reason also, it is advantageous to evaporate the smaller grains first because then 
oxides will be incorporated into the gas and the partial pressures of condensable elements 
will increase locally. This helps to make liquids stable at high temperature, and also 
suppresses the evaporation of silicate liquids which would have caused heavy isotope 
enrichment of resulting chondrules; such enrichment is not observed \citep{gal00}. 
\citet{ebe00} gives a full treatment of 
what dust enrichments are necessary to make FeO-bearing silicates and also to make 
chondrule liquids stable at low pressure relative to solids, as well as a discussion of 
dust-enrichment scenarios with high oxygen fugacities. The idea that the midplane is a 
better place to consider chondrule formation is supported again in this context.

The value of the gas density used in Figure \ref{fig4} is $n_g = 10^{12}$ cm$^{-3}$ 
corresponding to $f_{\Sigma} = 10^{-1}$. Since we expect the gas density in the disk 
to decrease to this level after a time of order a few million years, this 
argues for chondrule formation in a late, evolved disk, and  
is consistent with the age difference between CAIs (refractory Ca-Al-rich inclusions) 
and chondrules, estimated from the decay of $^{26}$Al (with a half-life of 0.7 Myr) 
to be 2--3 Myr \citep{hut94,rus96,swi96}.

\subsection{Current Sheets in the Disk Midplane}
\label{midplane}

From the results in the previous section, we conclude that in a standard model of the early 
solar nebula in which dust and gas are well-mixed, the disk will have $\zeta \approx 0.01$ 
everywhere and hence heating of dust grains by current sheets will not be effective.
This conclusion no longer holds in the late stage of the disk evolution. 
As argued by \citet{san00}, the bulk of the disk becomes unstable to the MRI, therefore 
susceptible to current sheet formation, for low enough values of the surface density 
factor $f_{\Sigma}$ or of the dust depletion factor $f_d$ because the 
recombination rate on grain surfaces decreases and thus $\chi_i$ rises except in a thin 
layer around the midplane. Current sheets capable of melting mm-sized dust grains form only 
when $n_g < 10^{12}$ cm$^{-3}$ and $\zeta$ is quite high, i.e. near the disk midplane. For 
these reasons, we henceforth 
consider a late stage of solar nebula evolution in which the gaseous disk has depleted 
by a factor of 10 ($f_{\Sigma} = 10^{-1}$) and settling of dust grains onto 
the midplane has proceeded significantly ($f_d \leq 10^{-2}$).

Let us first estimate the thickness of this dust-rich layer. We assume that $\zeta$ 
has increased from the well-mixed value of $10^{-2}$ to $50$. Taking 
into account the reduction in gas density, the vertical scale height of dust grains is 
expected to decrease by a factor of 500 compared to the gas scale height of 
0.1 AU at $R=2$ AU. Thus a thickness of $3 \times 10^4$ km is estimated.
This thin dust-rich layer around the midplane will be mostly magnetically dead because 
of the extremely high dust density there, and the resulting high recombination rate on grain 
surfaces \citep{san00}. Yet, the thin surfaces of this dust layer will be sufficiently 
ionized by cosmic rays and stellar X-ray radiation to support the MRI and generate 
current sheets, considering that there are now few dust 
grains suspended in the upper atmosphere. Current sheets that form in these thin 
layers will ionize nearby regions within the dust layer, into optical 
depth of order unity. Dust grains are heated after passing 
through these current sheets, sometimes up to their melting temperature ($T_{melt} \approx 
1600$ K) as shown in \S\ \ref{energy}, and then radiate strongly in the near-infrared, 
subsequently raising the temperature of dust grains in the surrounding environment. 
Collisions with hot dust grains may then ionize\footnote{The dominant ion species in 
protoplanetary disks are 
alkali metals (Na$^+$, K$^+$). Grains are dominant charge carriers at $T_g \lesssim 10^3$ 
K, but at higher $T_g$ desorption of charged ions from the surface of dust grains becomes 
effective \citep{ume83}. Thus, our neglect of the effect of charged dust grains is 
partially justified, at least inside current sheets.} the gas in an initially magnetically 
dead area above the critical value, $\chi_{crit}$. 

However, the integrated optical depth of the protosolar nebula at $R=2 \AU$ to the 
near-infrared radiation in the vertical direction is estimated to be $\tau = \kappa_{IR} 
\Sigma \approx 2 \times 10^2 $ \citep{dal01,chi01}. Even if we assume that some of the 
smallest, submicron-sized grains have been removed from the nebula with the gas, the 
magnetically active region affected by current sheets would still be only a few percent 
of the dust-rich 
layer. To melt the majority of the dust grains that reside in the midplane and form 
chondrules, the entire layer must be in motion so that all dust particles are exposed 
in the magnetically active surface layer at some point. 

When most of the solids have settled 
to the midplane, vertical shear between the gas and the differentially rotating, dense
particle layer can drive 
turbulence within a 10$^4$--10$^5$ km thick boundary layer \citep{cuz93,cuz96}. Although 
vertical velocity gradients vanish by symmetry very near the midplane, a turbulence model 
that self-consistently deals with generation, transport, and damping shows that the 
turbulence diffuses inward and persists even very near the midplane \citep[][and 
references therein]{cuz93}\footnote{But see \citet{sek98} and \citet{you02} for claims 
that turbulent mixing becomes ineffective if the dust-to-gas surface mass ratio is 
significantly enhanced.}. We take a typical velocity of $U \approx 10^2$ cm s$^{-1}$ 
[$\,$see 
Eqn. (65) of the previous reference$\,$] and the thickness of the dust layer, $L_d \approx 
3 \times 10^4$ km, and estimate the eddy turnover time to be 
$L_d/U \approx 1$ yr. Keeping in mind that the fraction of current sheets covering the dust 
surface is quite small ($l_{cs}/L_{BH} \approx 3 \times 10^{-5}$; see 
the next subsection), we expect an average dust grain to be exposed to current sheet 
heating about once every $3 \times 10^4$ yrs. After the brief heating, now spherical, mm-sized 
chondrules will settle down to the midplane, collecting rims of dust in the process. Such 
rims are observed \citep[][and references therein]{met96}.

\subsection{Near-Infrared Excesses}
\label{nearir}

The higher than predicted flux at 2--8 $\mum$ observed in the spectra of some young protoplanetary disks has been known for a decade \citep{har93,nat99,chi01}. Using a disk model that included an accounting of grain size distributions and grain compositions, and numerical solution of the equations of radiative and hydrostatic equilibrium under the two-layer approximation of CG97, \citet{chi01} rediscovered this in 4 of their 5 Herbig Ae and T Tauri stars, and found a relative dearth of emission shortward of 10$\mum$ arising partly from low absorption efficiencies of silicate grains. It is noteworthy that these sources also showed evidence for dust settling; parameters appropriate for gas and dust that are well-mixed in interstellar proportions did not fit the overall level of infrared excess at $\lambda \lesssim 100\mum$ for 4 of them. They interpreted this to mean that dust in disk surface layers has settled vertically towards the midplane.

\citet{dul01} proposed a disk model in which an inner rim located at the dust evaporation radius forms a vertical wall due to direct heating by the central star and, as a result, puffs up and intersects more starlight. Its reemitted flux produces a bump that peaks at 2--3$\mum$ in the SED. They note that this modification would be significant only for Herbig Ae stars. This mechanism works only if optical depth due to gas opacity in the inner hole is negligible, but in fact it is quite large, of order 10$^3$, when a standard surface mass density profile $\Sigma \propto R^{-3/2}$ is assumed, although the surface density of the inner disk is not constrained by the data \citep{nat01}.

We suggest an alternative mechanism by noting that current sheets at the dust melting temperature would radiate at just the right wavelengths to fill in the deficient flux at 2--8 $\mu$m, perhaps explaining the excess near-IR fluxes. A simple estimate can be made as follows. We take the mean separation between current sheets (produced by the Brandenburg-Zweibel mechanism) to be $L_{BH} = \lambda_{BH}/2$, roughly the distance between two adjacent magnetic nulls produced by the MRI. The total emission from dust grains inside current sheets is given by \citep{chi01}
\begin{eqnarray}
L_{\lambda} &\approx& \frac{l_{cs}}{L_{BH}} \left[ \, 8 \pi^2 \, \lambda \int_{1\AU}^{2\AU} 2\,B_{\lambda} (T_{melt}) \, R\,dR \, \right] \nonumber\\
&\approx& 10^{32} \left( \frac{\lambda_{\mu}^4 \, [\, \exp \, (9/\lambda_{\mu})-1 \, ]}{2 \times 10^3} \right)^{\!-1} \, \rm{erg \, s^{-1}} \, ,
\end{eqnarray}
for $n_g = 10^{12}$ cm$^{-3}$ at $R = 3 \AU$. Here, $\lambda_{\mu}$ is the wavelength of radiation in microns. Taking $l_{cs} \approx 10^2\km$ and $L_{BH} \approx 0.03\AU$, we obtain $l_{cs}/L_{BH} \approx 3 \times 10^{-5}$ for the fraction covered by current sheets in the region between 1 and 2 AU, the site of their formation in an evolved disk thinned by $f_{\Sigma} = 10^{-1}$ (see Figure \ref{fig1}b). This crude estimate for the excess near-IR emission from current sheets may partly explain the observed excesses in the range of 10$^{32}$--10$^{33}$ erg s$^{-1}$. Uncertainties in modeling of protoplanetary disks at present are admittedly large. Our mechanism involving current sheets could be taken as an example of producing near-IR excesses when magnetohydrodynamical effects are considered.

Ultimately, these ideas can be tested by searching for distinguishable observable signatures using techniques that have recently emerged. Interferometric imaging of LkH$\alpha$ in the H and K bands by \citet{tut01} has shown a central gap (or cavity) and a hot inner edge, observed at 3.4 AU from the central star when a distance of 160 pc to LkH$\alpha$ is assumed. Earlier theoretical predictions for its radius to fit the near-IR inflections in the SED \citep[$\sim10 R_*$;][]{hil92} are too small by around an order of magnitude. If we can divide the disk into several annular regions and obtain photometric measurements of each annulus, we suggest that there would be a local maximum of near-IR radiation between 1 and 2 AU, for systems like our primordial solar nebula (although for other young stellar objects, the exact position will depend on stellar and disk parameters and should be recalculated). In standard protoplanetary disk models such as CG97, the mentioned region reaches temperatures of only $100 -400$ K and is expected to emit poorly in the near-IR. In the near future, a multi-band, adaptive optics coronagraph under development at the American Museum of Natural History \citep{opp00} will be used to directly view protoplanetary disks and will present another possibility to test these theories.

\section{Caveats}
\label{caveats}

Our conclusions suffer from several caveats.
All models of the current sheets that have been done to date have assumed a steady 
state flow \citep[BZ94; ][]{bra95,zwe97,hei03}. However, the heating and ionization that occur in the current sheet 
are not accounted for self-consistently in these models. Both will increase the ion 
pressure in the center of the sheet, either saturating the singularity or even blowing the 
sheet apart. Also, we invoke turbulence induced by the MRI when calculating magnetic field 
fluctuations but ignore its effect on the evolution of current sheets, e.g. the 
converging motion of the neutrals in ambipolar diffusion. In our simplified chondrule 
formation model, we let gas and dust travel through a current sheet at $v_r \approx 1$ 
km s$^{-1}$, but it is not clear how well this approximates the actual condition near a 
current sheet or if a current sheet would survive in that environment. To study these 
effects in detail a dynamical model will be necessary. This requires modeling the interplay 
between at least four physical processes: pressure imbalance, ionization evolution, 
temperature evolution, and steepening of the magnetic field. These four processes are 
inherently connected and feed back on one another. That means they must be solved 
simultaneously in a self-consistent fashion. Radiative transfer and a power-law size 
distribution of dust grains must be incorporated in 
a subsequent step to understand chondrule formation in these dynamical models. The 
large range of timescales and length scales involved presents another challenge for 
modeling of this complicated set of processes.

\acknowledgments

We are grateful to Axel Brandenburg, Eugene Chiang, Fabian Heitsch, and Ellen Zweibel for 
helpful discussions, as well as an anonymous reviewer of an NSF grant proposal for pointing 
out the importance of high oxygen fugacities derived for chondrules. M. K. R. J. was supported 
by an AMNH Graduate Student Fellowship. M.-M. M. L. acknowledges support by NSF Career grant 
AST99-85392, NSF grant AST03-07793, and NASA ATP grant NAG5-10103. D. S. E. acknowledges 
support from NASA Cosmochemistry grant NAG5-12855.

\clearpage

\clearpage

\end{document}